\documentclass[twocolumn,iop]{openjournal}
\usepackage{graphicx}
\usepackage{txfonts}
\usepackage[colorlinks,allcolors=blue]{hyperref}

\usepackage{standalone}
\usepackage{tikz}
\usepackage{ifthen}
\usepackage{float}
\usetikzlibrary{matrix,calc}

\begin{document} 

\journalinfo{The Open Journal of Astrophysics}
\submitted{submitted February 2021, accepted March 2021}

   \title{Dwarfs from the Dark (Energy Survey): a machine learning approach to classify dwarf galaxies from multi-band images}
    \shorttitle{Dwarf from the Dark Energy Survey}

   \author{Oliver M\"uller$^{\star1}$   }
\author{ Eva Schnider$^2$}


\affiliation{$^{1}$Observatoire Astronomique de Strasbourg  (ObAS),
Universite de Strasbourg - CNRS, UMR 7550 Strasbourg, France}
\affiliation{$^{2}$Department of Biomedical Engineering, University of Basel, Allschwil, Switzerland}
\thanks{$^\star$ E-mail: \nolinkurl{oliver.muller@astro.unistra.fr}}


 
  \begin{abstract}
  Countless low-surface brightness objects -- including spiral galaxies, dwarf galaxies, and noise patterns -- have been detected in recent large surveys. Classically, astronomers visually inspect those detections to distinguish between real low-surface brightness galaxies and artefacts. Employing the Dark Energy Survey (DES) and machine learning techniques, \citet{2020ascl.soft11026T}  have shown how this task can be automatically performed by computers. Here, we build upon their pioneering work and further separate the detected low-surface brightness galaxies into spirals, dwarf ellipticals, and dwarf irregular galaxies. For this purpose we have manually classified 5567 detections from multi-band images from DES and searched for a neural network architecture capable of this task. Employing a hyperparameter search, we find a family of convolutional neural networks achieving similar results as with the manual classification, with an accuracy of 85\% for spiral galaxies, 94\% for dwarf ellipticals, and 52\% for dwarf irregulars. For dwarf irregulars -- due to their diversity in morphology -- the task is difficult for humans and machines alike. Our simple architecture shows that machine learning can reduce the workload of astronomers studying large data sets by orders of magnitudes, as will be available in the near future with missions such as Euclid. 
  \end{abstract}

   \keywords{  Galaxies: dwarf; Galaxies: spiral; Galaxies: elliptical and lenticular, cD; Galaxies: irregular;  Methods: data analysis              }

   \maketitle
%

\section{Introduction}
We are living in a golden age to explore the low surface brightness Universe. Current instrumentation, paired with cutting-edge computing facilities, allows the study of the Universe in unprecedented detail. For our own Milky Way system, large surveys like the Sloan Digital Sky Survey (SDSS, \citealt{2000AJ....120.1579Y}) or the Dark Energy Survey (DES, \citealt{2018ApJS..239...18A}), have increased the number of faint dwarf galaxies by several magnitudes \citep[e.g.,][]{2005ApJ...626L..85W,2010ApJ...712L.103B,2015ApJ...804L..44K,2015ApJ...813..109D}. With this knowledge came new opportunities to probe our understanding of gravity \citep{2010MNRAS.406.1220W,2016ApJ...832L...8M,2018MNRAS.480..473F}, the role of baryons \citep{2012MNRAS.421.3464P,2014ApJ...795L...5S,2018A&A...616A..96R}, and cosmology \citep{2010A&A...523A..32K,2011MNRAS.415L..40B,2012MNRAS.423.1109P}. 

Outside of our own Milky Way system, the search for and study of these elusive dwarf galaxies has mainly relied on targeted observations of selected galaxy groups and clusters \citep[e.g.,][]{2009ApJ...705..758M,2009AJ....137.3009C,2014ApJ...787L..37M,2014ApJ...793L...7S,2015A&A...583A..79M,2016ApJ...828L...5C,2017ApJ...848...19P,2017A&A...608A.142V,2018ApJ...855..142E,2018ApJ...863..152S,2018ApJ...868...96C,2018ApJ...867L..15T,2020ApJ...891..144C,2020MNRAS.491.1901H,2020A&A...642A..48I}. Both automatic source extraction and visual scans of the images -- or a mix of both -- are employed to search for dwarf galaxies, even today \citep[e.g.,][]{2020MNRAS.491.1901H,2020A&A...644A..91M}. These are tedious and time consuming tasks and need some expert knowledge to correctly identify the dwarf galaxy candidates. 

Follow-up observations of many dwarf galaxy candidates from some of these surveys with better instruments have revealed an unavoidable contamination of false detections, i.e. by confusion with background galaxies, foreground galactic cirrus, or instrumental noise \citep{2013AJ....146..126C,2016ApJ...833..168M,MuellerTRGB2018,MuellerTRGB2019,2019ApJ...885..153B}. A prime example is the galaxy Cen\,8/KK\,198, which has been regarded as dwarf galaxy associated with the Centaurus group for more than two decades \citep{1996PASA...13T.278C,2000AJ....119..593J,2013AJ....145..101K,2017A&A...597A...7M}, until VLT observations have uncovered it as a low-surface brightness spiral galaxy \citep{MuellerTRGB2019,2021A&A...645A..92M}. Once high-resolution imaging was available to study the morphology of the galaxy in detail, the spiral pattern in the older imaging becomes quite apparent, in other words, this spiral galaxy could have been spotted as an interloper in the dwarf galaxy catalogs before the costly follow-up observations.

With upcoming surveys like LSST or Euclid, we will face a surge of low-surface brightness objects. If we want to study the dwarf galaxy regime with these immense data sets, we need novel approaches -- hand-on categorization of all the faint sources will be too time-consuming for humans to do. Fortunately, we can teach computers to perform such tasks. With the advent of deep neural networks, computer vision has become a solvable task and has found a broad range of usages, from recognizing handwritten letters \citep{2015arXiv151108458O}, human faces \citep{Huang2012a}, individual  bones \citep{10.1007/978-3-030-59861-7_5}, animals \citep{2014arXiv1409.0575R}, and plants \citep{Goau2017PlantIB}, to even being able to fake pieces of arts \citep{Gatys2015ANA}, pizza \citep{Papadopoulos2019Pizza}, or videos \citep{korshunov2018deepfakes}. 

The use of neural networks to classify astronomical images and a discussion of their potential role in handling large digital suveys dates back to the early nineties \citep{storrie1992morphological, bertin1994classification}. More recently, \citet{2020ascl.soft11026T} have used a simple three layered convolutional neural network (CNN) to separate artefacts from low-surface brightness objects they previously found  in data from the DES \citep{2021ApJS..252...18T} to a 90\% accuracy. Even more encouraging is the fact that transfer learning -- meaning to use the pre-trained network on a different data set -- seems to be doable with minimal effort of re-training the model with a subsample from the new data set, in the \citet{2020ascl.soft11026T} case data from the Hyper Suprime Cam survey. 

In this work, we build upon the work of \citet{2020ascl.soft11026T} do further distinguish their low-surface brightness objects into spiral and dwarf galaxies, employing machine learning. To do so, we first manually classify  5567 objects, then we find the best set of hyperparameters to define our neural network, and then finally train this network. Before jumping into the topic of finding a suitable architecture, we will give a brief introduction to neural networks, focusing on the techniques we are going to apply. The code can be found on GitLab\footnote{https://gitlab.com/VoltarCH/deeplearning\_des}.

\section{Neural Networks}

Much research has been conducted in the past years to find the most promising machine learning methods for classification of images \citep{krizhevsky2012imagenet, lecun1989backpropagation, simonyan2014very}. While there is a plethora of possible architectures, many supervised classification tasks work well using comparatively simple convolutional networks. Specifically variations of LeNet \citep{lecun1989backpropagation} in combination with modern optimisation techniques are still wide-spread \citep{ozyurt2019brain, sardogan2018plant}. A sequence of convolutional layers with down-pooling is followed by a few fully-connected layers, ending with a classification layer consisting of as many nodes as classes of interest.

Convolutions have been used in digital image processing for a long time, also outside the context of neural networks \citep{1977AASPB..16...10M,keys1981cubic, perona1990scale}. Every pixel of the output image consists of the dot-product between the surrounding pixels of the input image and the entries of the convolutional kernel. Depending on their values kernels can serve many purposes, such as edge detection, blurring, or sharpening of an image. In convolutional neural networks, the values of the convolutional kernels are incrementally learnt, instead of hand-crafted. To every input layer, multiple convolutional kernels can be learnt and applied. The resulting outputs of those individual kernels are called channels. The number of channels thus directly corresponds to the number of learnt kernels.

In fully connected layers, every node $n_{out}^j$ is the weighted sum of all nodes of the previous layers, plus an offset: $n_{out}^i = \sum_i w^{i,j}*n_{in}^{j}+b^{j}$. The weights $w$ and biases $b$ of the whole network are trainable parameters and are iteratively updated while training the network. As the name suggest, in fully connected layers every node of a layer is linked to every node of the preceding layer. In contrast, the local dot-product of convolutional layers only links nodes that are spatially close.

Since even an elaborate chain of linear functions stays a linear function, activation functions are added to the layers of a neural network. These simple non-linear functions allow for the possibility that the output of the neural network does not need to linearly depend on the input \citep{hinton2012deep}.

Pooling layers are used to down-sample the input. They work similar to convolutional layers, by applying operations to a small region of input at once. In contrast to convolutional layers, the kernel of pooling layers is not trained but fixed up-front. Two popular examples include a kernel that chooses the maximum value of the input region, or the mean value of the input region. The downsampling effect is achieved by applying the kernel such that the input regions don't overlap. That way, a 2x2 pool kernel reduces the spatial size of the output by a factor of 2 per spatial dimension.

The trainable parameters of a neural network are iteratively updated as part of an optimisation process. Images are passed through the network and the output -- the predicted class -- is compared to the true class of the image. A common loss function to quantify the agreement is the cross-entropy loss. Using the chain rule, the gradient of the loss with respect to every single trainable parameter can be computed. This in turn allows for the use of gradient-based optimisers. A lot of research has been conducted to find first-order-gradient based optimisers that show good performance in training neural networks, with a certain robustness against ending up in local optima or getting stuck in ravines of the search space \citep{zeiler2012adadelta, kingma2014adam, qian1999momentum}.

Modern classification network architectures tend be very deep, i.e. consist of large numbers of convolutional layers \citep{szegedy2015going, krizhevsky2012imagenet, he2016deep}. Due to the resulting high number of trainable parameters and to prevent overfitting, they are generally trained on tens of thousands of natural images \citep{krishna2019deep}. Training these kinds of architectures from scratch costs a lot in terms of computational memory and time. A popular approach is therefore to leverage the power of transfer learning for new tasks \citep{pan2009survey}. In this case, network parameters are not initialized using random functions, but using the final parameter values of an openly accessible fully trained network. It is then possible to update the parameters during future training on the new task. To extend the network's predictive capabilities to new classes -- unseen during the initial training phase -- the very last classification layer is dropped and replaced by one which consists of the new number of output classes. While this approach reduces the number of data needed to learn a new task on very deep networks, the computational requirements remain huge, due to the millions of trainable parameters. This problem can be alleviated by fixing all parameters but the ones in the classification layers. In this way the number of trainable parameters is reduced tremendously, which speeds up the training considerably and allows the use more accessible hardware. However, there are also drawbacks to keeping most of the layers fixed. The inner layers learn to encode images in an abstract space which facilitates the classification. Since this internal encoding was learnt during the initial training using natural images, other types of images may not be represented well using this encoding. Therefore, the success of transfer learning from state-of-the art natural image classification networks for types of imaging such as histology, medical imaging, or astronomical images depends on the choice of network, data set size and task \citep{zhang2016automatic,cha2017bladder,martinazzo2020deep}.

\section{Data}

\begin{figure*}[ht]
    \centering
    \includegraphics[width=0.24\linewidth]{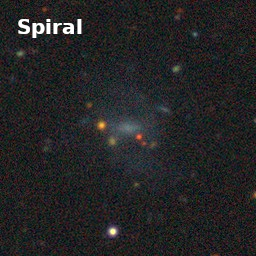}
    \includegraphics[width=0.24\linewidth]{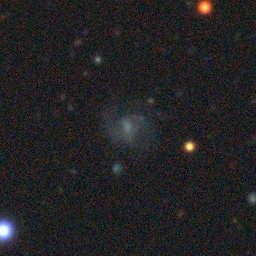}
    \includegraphics[width=0.24\linewidth]{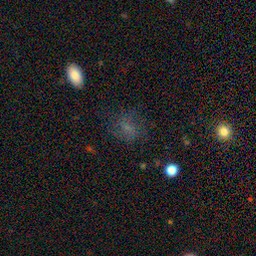}
    \includegraphics[width=0.24\linewidth]{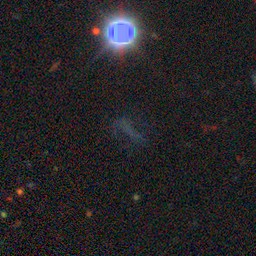}\\
        \includegraphics[width=0.24\linewidth]{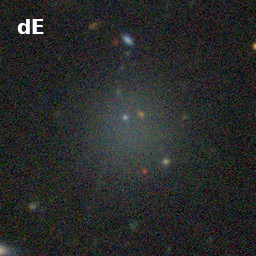}
    \includegraphics[width=0.24\linewidth]{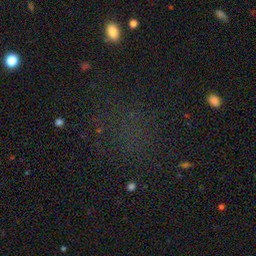}
    \includegraphics[width=0.24\linewidth]{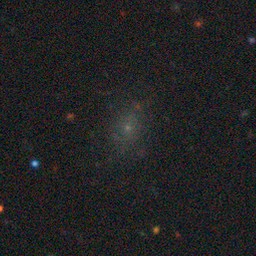}
    \includegraphics[width=0.24\linewidth]{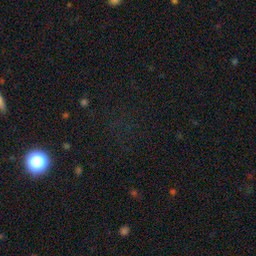}\\
        \includegraphics[width=0.24\linewidth]{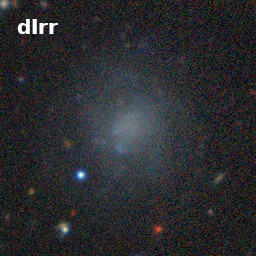}
    \includegraphics[width=0.24\linewidth]{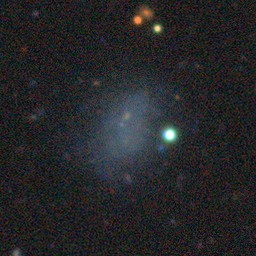}
    \includegraphics[width=0.24\linewidth]{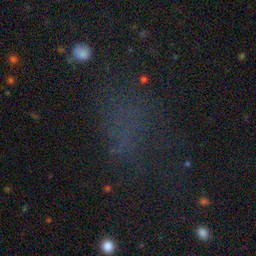}
    \includegraphics[width=0.24\linewidth]{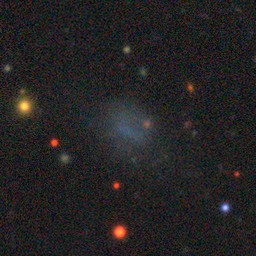}
    \caption{Typical objects corresponding to the three classes manually labeled in this work: spirals (top row), dwarf ellipticals (middle row), and dwarf irregulars (bottom row). The images have a side of 256\,px.}
    \label{fig:classes}
\end{figure*}

\begin{table*}[ht]
\caption{Hyperparameters, their optimization search space and optimal value.}   
\label{table:hyper}      
\centering          
\begin{tabular}{l l l l}     
\hline\hline       
Name & search space & best value & description\\
\hline                    
$n_{layers, conv}$ & $\left[1, 5 \right]$& 5 & number of convolutional layers  \\
$n_{channel}$ & $\left[2, 200 \right]_2$& 40  & number of initial channels in the first convolutional layer, with a step size of 2 \\
$n_{nodes}$ & $\left[10, 200 \right]$ & 108 & number of nodes in the second fully connected layer\\
$\delta$ & $\left[0, 0. \right]$& 0.84 & negative slope of the Leaky ReLU activation functions \\
$p_{dropout}$ & $\left[0, 0.9 \right] $& 0.28 & dropout rate for regularization\\
image size& $\left[32, 256 \right]_{32}$  & 96 & size of one side of the input image, , with a step size of 32\\
 optimizer  & Adam, SGD, RMSprop & RMSprop & the optimizer used to update trainable parameters \\
learning rate &  {[}$10^{-5}, 10^{-1}${]}& $1.84\times 10^{-5}$ & learning rate used by the optimizer \\
momentum & $\left[0, 0.99\right]$& 0.78 & momentum used by the optimizer \\
weight decay &  $\left[0, 0.2\right]$ & 0.003 & weight decay used by the optimizer \\
batch size &  $\left[2, 128 \right]_2$& 2 & the size of the batch for the training, with a step size of 2 \\
\hline                  
\end{tabular}
\end{table*}


\citet{2020ascl.soft11026T} selected a set of 20'000 low-surface brightness objects detected in DES  \citep{2021ApJS..252...18T} and predicted them to be real low-surface brightness objects (in contrast to artefacts) by the use of machine learning techniques.
We downloaded the centered images of these objects from the DES archive using a modified script provided in the git project\footnote{https://github.com/dtanoglidis/DeepShadows} of \citet{2020ascl.soft11026T}. The jpeg images have the dimensions of $256\times256$\,px, corresponding to a field of view of $1.1\times1.1$\,arcmin$^2$, and have three channels, i.e. the standard RGB channels (not to be confused with the sloan $r$ and $g$ and Johnson/Bessell $B$ filters). The pixel values span from 0 to 255, with 0 corresponding to no intensity in the pixel.

 Because many of these detections are small background spiral galaxies, we have made a size cut of 3.3\,px on their table column named flux\_radius\_r, which was measured \citep{2021ApJS..252...18T} with Source Extractor \citep{1996A&AS..117..393B}. This left us with 5567 low-surface brightness objects.  Assuming a distance of 50\,Mpc, this cut  translates into a radius of 200\,pc, corresponding to a typical effective radius for a faint dwarf \citep{MuellerTRGB2019}. The distance of 50\,Mpc was chosen to mock the selection of galaxies from the MATLAS survey \citep{2020arXiv200713874D}.

To further increase the sample size during the training and validation of our neural networks, we used standard data augmentation techniques used in machine learning. Namely, each image can be flipped (i.e. mirrored) and rotated (by 90, 180, or 270 degrees). These data augmentations increase the number of images used in this work by a factor 8 to 44536. We have decided not to use translations or rotations which would need interpolation of the pixel values. Such transformations could lead to unexpected behaviors of the neural networks, but it could be interesting for future studies to include them as well.

We will split the sample into three different sets, a training, a validation, and a test set, which are randomly drawn from the total sample of galaxies. The training set consists of 70\%, the validation set of 10\%, and the test set of 20\% of the data. As the name suggests, the training set will be used for the training of the network. The validation set will be used for the hyperparameter search, and the test set for the final evaluation of the trained network.

\section{Manual Classification}

To train deep neural networks, it is necessary to provide many labelled samples of the classes we want them to be able to distinguish. Because the data set provided by \citet{2020ascl.soft11026T}  is not morphologically labelled, we had to perform the classification by ourselves. For this purpose, we have used Pidgey\footnote{https://github.com/wbwvos/pidgey.}, a simple Jupyter notebook tool to annotate images. We have created three classes, these are i) spiral galaxies, ii) dwarf ellipticals, and iii) dwarf irregulars. This follows the classification scheme implemented in \citet{2020MNRAS.491.1901H}. We have classified an object considering following typical features of each class: 

\begin{itemize}
    \item {spiral:} indications of a bar; spiral arms; extended nucleus;  compact shape; high ellipticity indicating viewing the spiral edge-on, sharp edges;  strong color gradient from red in the center to blue in the outskirts. 
    \item dwarf elliptical: smooth profile; extended; diffuse; no irregular features.
    \item dwarf irregular: mostly smooth profile; extended; irregular features (e.g. tidal tails, star forming regions),  which are not spiral arms.
\end{itemize}
It was not always straight forward to classify an object, especially between small dwarf ellipticals and spirals. If an object appears smooth, but very small, it is likely an unresolved spiral and therefore was classified as spiral galaxy. Of course this could falsely label some small dwarf elliptical as spiral galaxies. Also the differences between disturbed spiral galaxies and dwarf irregular are sometimes difficult to tell. Biases in human classifications are therefore unavoidable. To  improve the reliability of human classification, we have repeated the classification three times and made a majority vote, meaning that a label had to be picked at least two times to count. This gives also a base-line for the variability of the classification coming from the same person.

In total, we have labeled 2925 objects to be spiral galaxies, 2037 to be dwarf ellipticals, and 605 to be dwarf irregulars. For those, 2431 spirals (83.1\%), 1658 dwarf ellipticals (81.4\%), and 376 dwarf irregulars (62.1\%) have received three times the same vote, meaning that we were confident about the label. Uncertainties in labelling will have two main sources: one is coming from assigning them a wrong label (i.e. clicking on the wrong button), the other is coming from the true uncertainty about the morphology of the object. The latter is difficult to assess, because we do not know the ground truth of these objects and therefore can't directly quantify it.

\begin{figure*}[ht]
    \centering
    \includegraphics[width=\linewidth]{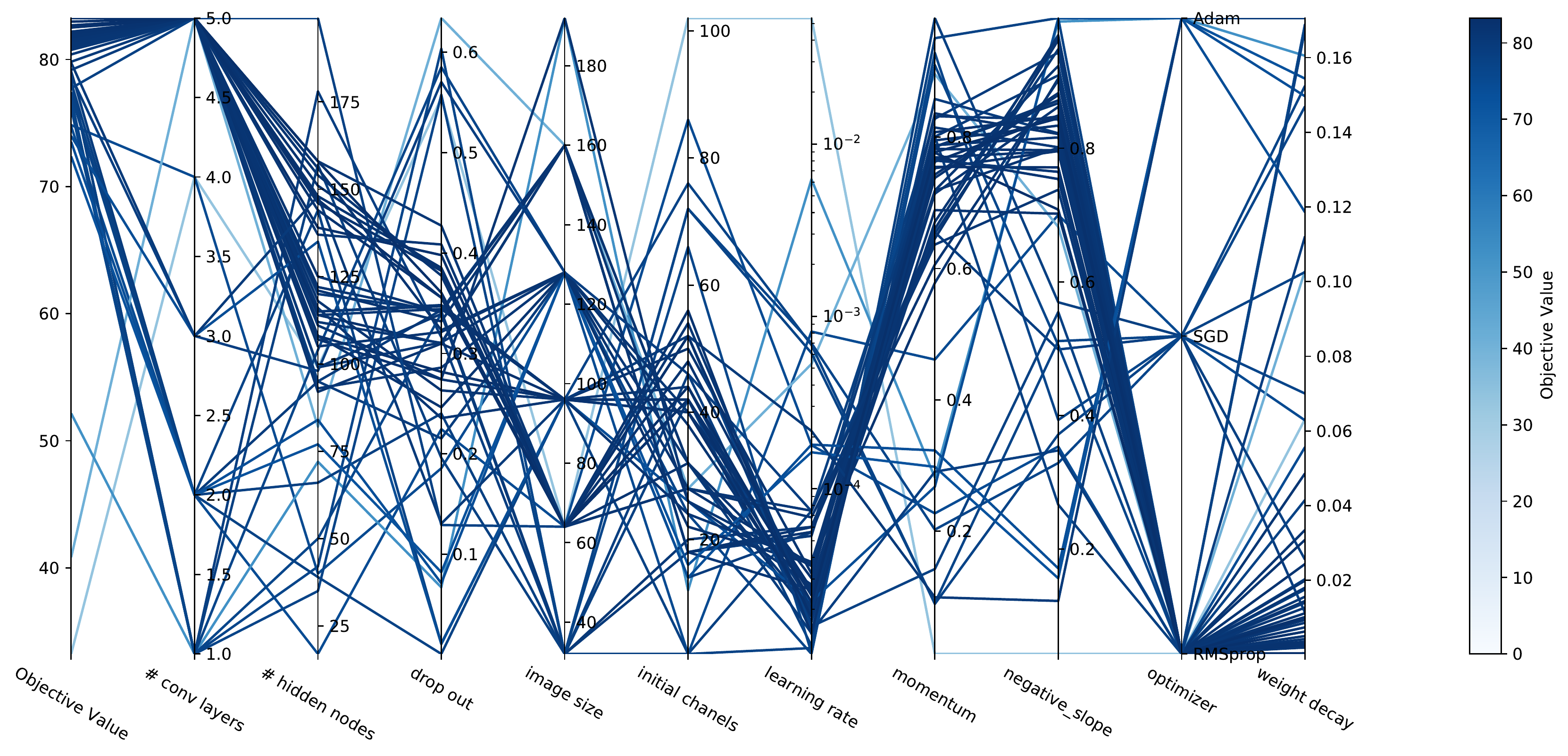}
    \caption{Parallel coordinate plot, indicating which combination of hyperparameters resulted in which performance, measured as objective value -- in our case the validation accuracy. }
    \label{fig:parallel_coordinates}
\end{figure*}

\section{Hyperparameter Search and Training}



\begin{figure}[ht]
    \centering
    \includegraphics[width=\linewidth]{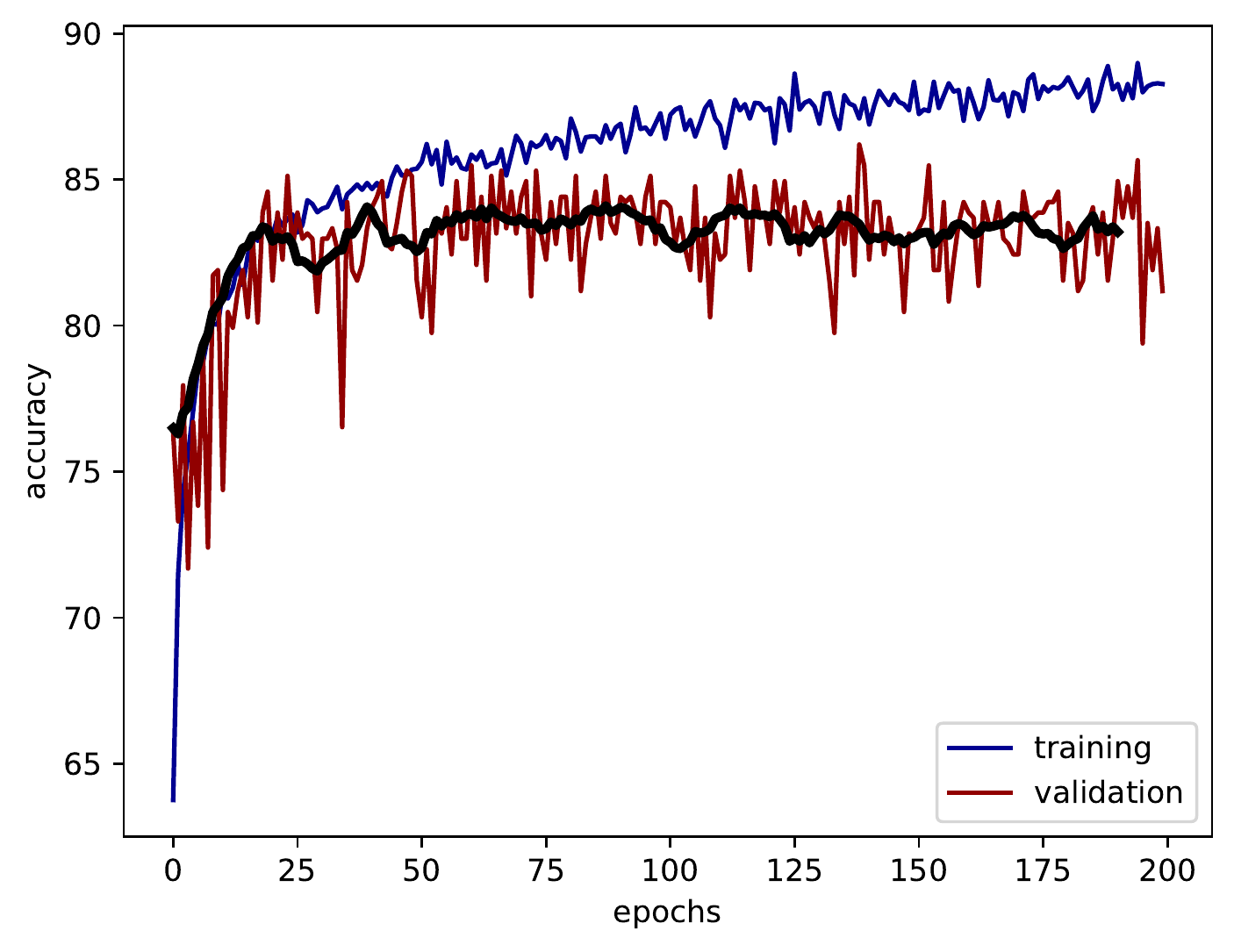}
    \caption{Training (blue line) and validation (red line) accuracies as a function of the training epoch. The running average of 10 epochs of the validation accuracy is indicated as thick black line. }
    \label{fig:validation}
\end{figure}

The search for the best set of hyperparameters is one of the most cumbersome tasks in any machine learning project, due to their high dimensionality and non-trivial inter-dependencies between them. In machine learning, hyperparameters describe values which need to be decided \textit{a priori} to the actual training of the neural network. The choice of these parameters can come from previous estimations of similar problems. Better though is the selection of the hyperparameters by some kind of pre-defined rules. The search for the best set of hyperparameters is an active research field, which continues to deliver novel approaches for this task. In our work, we use the Optuna framework \citep{akiba2019optuna}, which uses a multivariant Tree-structured Parzen Estimator (TPE, \citealt{bergstra2011algorithms}). The TPE estimates the densities of  good and bad hyperparameters. From this, it samples the most promising hyperparameters. Furthermore, Optuna uses pruning, which allows for early-stopping of bad sets of hyperparameters, speeding up the process of the hyperparameter search. For the pruning, we use Optuna's implementation of Hyperband \citep{li2017hyperband}.

The define-by-run principle implemented in Optuna allows to dynamically construct an optimal neural network from the range of hyperparameters and a pre-defined heuristic. Here, we use the validation accuracy as such heuristic. We start with a standard layout for CNNs, combining a convolution with a kernel size of 5 with a max-pooling (halving the spatial sizes). The number of these layers $n_{layers,conv}$ is a hyperparameter. The number of feature maps (also called channels) is doubled in each subsequent layer. The initial number of feature maps $n_{channel}$ is another hyperparameter. 
After these convolutions, a series of fully connected layers reduces the features, up until the final classification layer. We use two fully connected layers. The first fully connected layer has $n_{nodes}$ nodes, which is a hyperparameter, the last one has the number of labels, i.e. 3 nodes. 

We employ Leaky ReLUs as activation functions, with the negative slope $\delta$ being  another hyperparameter.    Throughout the network, dropout is applied as regularization, with the dropout rate $p_{dropout}$ being another hyperparameter. Further hyperparameters are coming from the optimizer: the choice of the optimizer itself, its learning rate, momentum, and weight decay. Finally, the image size is one more hyperparameter, with a step size of 32\,px per side, up to 256\,px. However, for the larger image sizes we sometimes run out of memory on our local machine, depending on the other hyperparameters. 

In total we have a set of 11 hyperparameters. They are presented in Table\,\ref{table:hyper}. We have used Optuna to search for the best combination of these hyperparameters using 500 trials with a maximum of 30 epochs in each run, using a cross entropy loss function.  The best set of hyperparameters is presented in Table\,\ref{table:hyper}. The networks themselves were implemented in PyTorch 1.7.0 \citep{NEURIPS2019_9015} and the experiments and training were conducted on a single nvidia GeForce RTX 2070 GPU with 8 GB of RAM and CUDA 11.0.


Now that we have the best set of hyperparameters estimated, we only need to decide for the number of epochs the network will get trained. For this, we again use the validation set, but with the hyperparameters fixed. We train our network for 200 epochs to see where overfitting starts to kick in. For this we measure the training and validation accuracy as a function of the epoch, see Fig.\,\ref{fig:validation}. While the network gets better on the training set the longer it trains, the validation accuracy starts to stay the same after $\approx$60 epochs, indicating that the networks starts to overfit. We therefore stop the training for the final evaluation of the network at 60 epochs.

To compare our CNN to a more sophisticated, pre-trained network from the literature, we also study the capabilities of ResNet50 \citep{xie2017aggregated} for our task at hand. Because it has too many parameters to fully train on our machine {-- the network contains 50 layers --}, we only re-train the last fully connected layer, which is just before the classification layer. ResNet50 requires as input an image with at least a dimension of $224 \times 224$\,px, therefore we simply use the original 256\,px images.  
We searched for the best set of hyperparameters with Optuna, but leaving out architectural choices and only searching for the optimal batch size, optimizer, learning rate,  momentum, and weight decay. They are: 44 for the batch size, SGD for the optimizer, $6.3\times10^4$ for the learning rate, 0.92 for the momentum, and 0.095 for the weight decay.  Similarly as before, we also searched for the epoch where overfitting starts to become relevant. This occurs at 50 epochs. Finally, we train and evaluate on the same training and test sets as before.

\section{Results and Discussion}

\subsection{Suitable network architectures}

\begin{figure*}[ht]
    \centering
    \includegraphics[width=\linewidth]{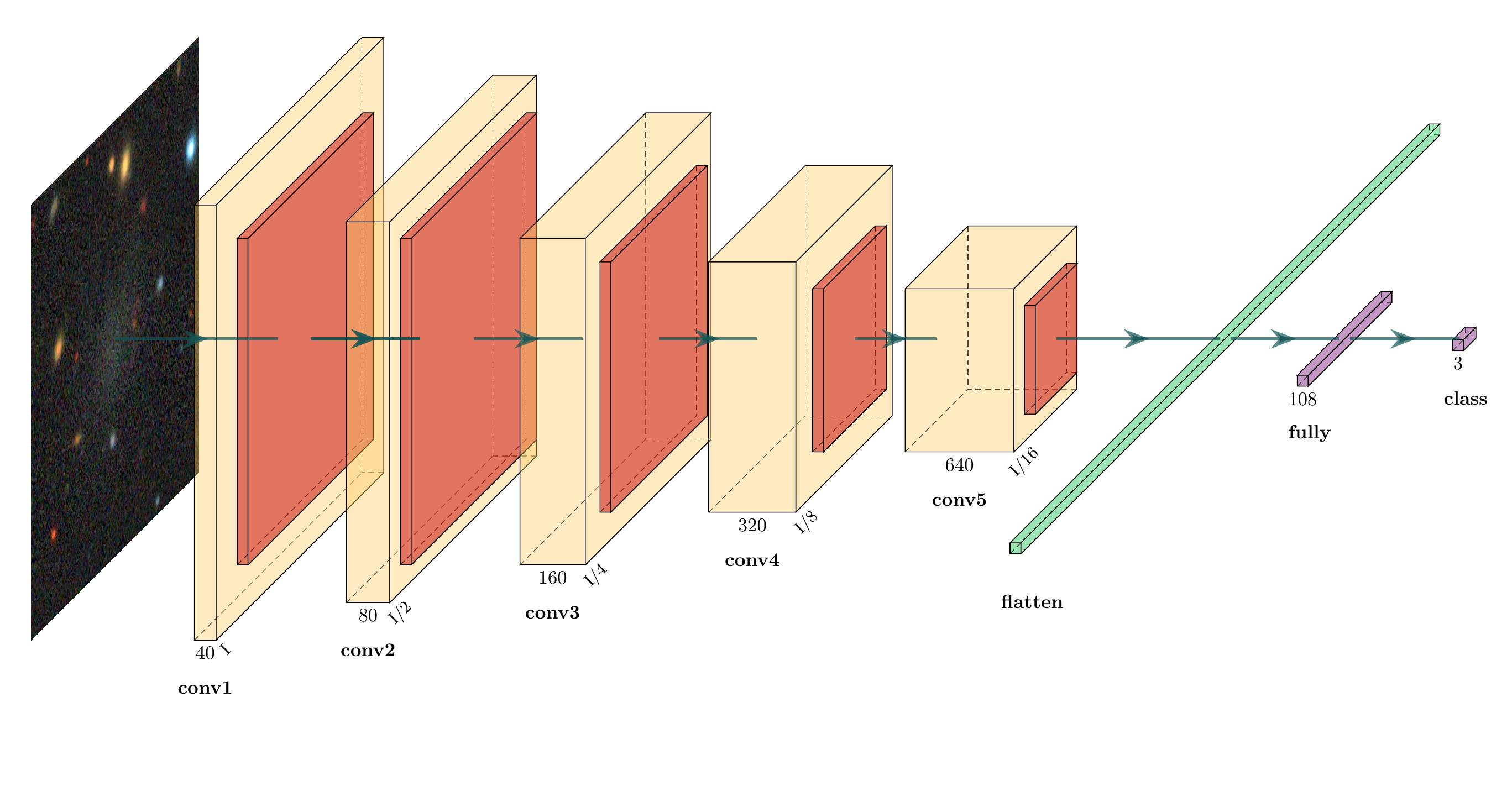}
    \caption{The optimal network as estimated from our hyperparameter search. It contains a sequence of convolutional (yellow) plus max pooling (red) layers, followed by a flattening layer (green), which maps all the nodes of the last convolution to a a linear layer, and then  a sequence of two fully connected layers (violet).}
    \label{fig:nn}
\end{figure*}

The best set of hyperparameters from our optimization is presented in Table\,\ref{table:hyper}. However, this set is not the only architecture yielding good results.
In Fig\,\ref{fig:parallel_coordinates} we present the performance for different combinations of the hyperparameters in a so-called parallel coordinate plot. This figure visualises the impact of the different combinations of hyperparameters and indicates whether certain hyperparameters must be tuned to achieve a good result, or if there are multiple ways to achieve similar results. The latter is the case for many of the hyperparameters, meaning that there is no distinct architecture superior to others.   The best results (i.e. highest objective values) are achieved with a family of network having similar, albeit not the same, hyperparameters, which are actually close to the proposed standard values in PyTorch, with some exceptions. Most surprisingly, the RMSprop optimizer yields the best results, although Adam is the standard for this task.
We present a graphical representation of the architecture achieving the highest   validation accuracy during the trials in Fig.\,\ref{fig:nn}.

\subsection{Evaluation of the neural network}

\begin{figure}[ht]
    \centering
 
\def\myConfMat{{
{499,  17,  32},  
{  68,391,22},  
{  20,  6,58},  
}}

\def\classNames{{"S","dE","dIrr"}} 

\def\numClasses{3} 

\def\myScale{1.5} 
\begin{tikzpicture}[
    scale = \myScale,
    ]

\tikzset{vertical label/.style={rotate=90,anchor=east}}   
\tikzset{diagonal label/.style={rotate=45,anchor=north east}}

\foreach \y in {1,...,\numClasses} 
{
    \node [anchor=east] at (0.4,-\y) {\pgfmathparse{\classNames[\y-1]}\pgfmathresult}; 
    
    \foreach \x in {1,...,\numClasses}  
    {
    \def\totSamples{0}
    \foreach \ll in {1,...,\numClasses}
    {
        \pgfmathparse{\myConfMat[\ll-1][\x-1]}   
        \xdef\totSamples{\totSamples+\pgfmathresult} 
    }
    \pgfmathparse{\totSamples} \xdef\totSamples{\pgfmathresult}  
    
    \begin{scope}[shift={(\x,-\y)}]
        \def\mVal{\myConfMat[\y-1][\x-1]} 
        \pgfmathtruncatemacro{\r}{\mVal}   %
        \pgfmathtruncatemacro{\p}{round(\r/\totSamples*100)}
        \coordinate (C) at (0,0);
        \ifthenelse{\p<50}{\def\txtcol{black}}{\def\txtcol{white}} 
        \node[
            draw,                 
            text=\txtcol,         
            align=center,         
            fill=black!\p,        
            minimum size=\myScale*10mm,    
            inner sep=0,          
            ] (C) {\r\\\p\%};     
        \ifthenelse{\y=\numClasses}{
        \node [] at ($(C)-(0,0.75)$) 
        {\pgfmathparse{\classNames[\x-1]}\pgfmathresult};}{}
    \end{scope}
    }
}
\coordinate (yaxis) at (-0.3,0.5-\numClasses/2);  
\coordinate (xaxis) at (0.5+\numClasses/2, -\numClasses-1.25); 
\node [vertical label] at (yaxis) {Predicted Class};
\node []               at (xaxis) {Actual Class};
\end{tikzpicture}
\caption{Confusion matrix, where the x-axis corresponds to the visually assigned label and the y-axis to the predicted label from the neural network.}   
\label{table:confusion}   
\end{figure}

The final evaluation of the network is done with the test set, which the network has not seen up to this point. The network was trained  during 60 epochs, as evaluated in our overfitting experiment. The accuracy on this test set gives us the performance. After training the network for 60 epochs, it is able to correctly classify 85\% of the spiral galaxies, 94\% of the dwarf ellipticals, and 52\% of the dwarf irregulars. In total it achieves an accuracy of 85\%. 

Has the network learned something? If this is the case, it must perform better than random. Because the three labels are not equally distributed, they have following probabilities to occur by chance: for spiral galaxies it is 52\%, for dwarf ellipticals it is 37\%, and for dwarf irregulars its 10\%. This indeed shows that for each class, the network was able to learn from the training data and use the generalization to classify the objects.

The confusion matrix of the classification is shown in Fig.\,\ref{table:confusion}. This matrix shows on the diagonal the correctly classified objects (i.e. the predicted label is equal to the ground truth set by our manual classification). The off-diagonal entries represent the confusion between the classes. If we look at the objects with spiral galaxies as ground truth, 85\% were correctly classified and the main confusion happened with the dwarf ellipticals (12\%). Only 3\% of the `real' spiral galaxies were classified as dwarf irregulars. For the dwarf ellipticals, 94\% of the test objects were correctly classified and only 4\% were confused with spirals and 1\% confused with dwarf irregulars. These two labels indeed yield good results with minimal confusion. However, for dwarf irregulars a different picture emerges. While 52\% of the test objects were correctly labelled, there was heavy confusion with both spirals (29\%) and dwarf ellipticals (20\%). This shows that there is ample room for improvement for dwarf irregulars.

\subsection{Human vs. Machine Performance}
\begin{figure}[ht]
    \centering
    \includegraphics[width=0.49\linewidth]{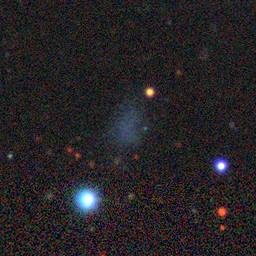}
    \includegraphics[width=0.49\linewidth]{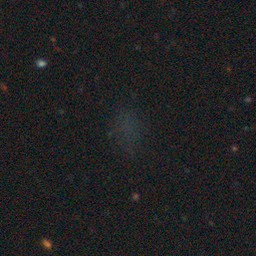}
    \caption{The galaxies which were labeled differently during the different manual classification runs. Left is a galaxy with a majority vote labelling it as dwarf irregular (but has one out of three votes as dwarf elliptical ),  and right a galaxy finally labeled as dwarf elliptical (but has one vote as dwarf irregular).}
    \label{fig:misclassified_human}
\end{figure}

Can a machine achieve similar results as a human? In our case, a direct comparison is tricky, because the ground truth -- i.e. what morphological type the object belongs to -- is still unknown. The repeated manual classification of the objects, however, may be used as a proxy for the accuracy of human classification. If the object is classified in each run the same way, this indicates that we are certain about the type of the object, if on the other hand the classification changes, this indicates uncertainty. So how uncertain were our manual classifications? For this we counted the fraction of labels which we classified the same each time. For the spiral galaxies, this happened in 83\% of the cases, for the dwarf ellipticals in 81\% of the cases, and for the dwarf irregulars in 62\% of the cases, highlighting how difficult dwarf irregulars are. If we rather ask the question if an object is a spiral galaxy or a dwarf galaxy (i.e. we do not care whether an object is a dwarf irregular or dwarf elliptical), the later number increases to 86\%. This shows that the confusion of dwarf irregular happens mainly with dwarf ellipticals. Two such cases are presented in Fig\,\ref{fig:misclassified_human}, where the vote was 2/1 in favor of them being a dwarf irregular or dwarf elliptical, respectively.

The uncertainty of human classification can now be used as a benchmark for the neural network. If we want to replace human interception with a machine, it should at least be as good as a human, if not better. So, what accuracy does our best network achieve? For spiral galaxies it achieves an accuracy of 85\% (vs. 83\% from the manual labelling uncertainty), for  dwarf ellipticals an accuracy of  94\% (vs. 81\%), and for dwarf irregulars 52\% (vs. 62\%). This is well compatible with our manual classification and shows that the network is capable of performing the job as good as a human expert. For dwarf ellipticals the neural network is even better, however, for dwarf irregular it performs worse. This is expected, as the dwarf irregulars have the least amount of training data and are the hardest to classify.

\subsection{Comparison to ResNet50}
We have shown that our simple CNN performs quite well given the data set at hand. So how is the performance of a more sophisticated network, namely ResNet50, for which the last fully connected layer has been retrained on our data set? Our adopted ResNet50 achieves a total test accuracy of 65\%, with 91\% of the spirals correctly classified,  42\% of the dwarf ellipticals, and 0\% of the dwarf irregulars. While it outperforms the classification of the spirals, in general it performs worse than our simple CNN. What is likely to happen is that it prefers to label spirals because they are the most common type in our data set, as well as have some distinct features (like rotation), which do occur in natural images. But the generalization power of ResNet50 {for the purpose of low-surface brightness classification  with our limited resources to re-train it} is less strong than a more simple convolutional network, like the one we found here. {This is probably due to the fact that ResNet50 was trained on images from the  ImageNet dataset \citep{deng2009imagenet}, which  highly differ from astronomical images in terms of dynamical range and features. }

\section{Summary and Conclusions}
In this work, we have shown how a neural network can be designed to distinguish  from multiband images between dwarf elliptical, dwarf irregular, and spiral galaxies to a similar accuracy as achieved by manual intervention. For this purpose, we have first classified  5567 galaxies detected in the Dark Energy Survey. Then, we used an optimization scheme to search for suitable architectures and hyperparameters. Our hyperparameter search revealed that a convolutional neural network with five convolutional layers and two fully connected layers with hyperparameters close to standard values yields the best results.

This network  was trained and evaluated on a separate set of data. For spiral galaxies, the network achieves an accuracy of 85\%, for dwarf ellipticals an accuracy of 94\% and for dwarf irregulars an accuracy of 52\%. Compared to the uncertainty of human classification, which we measured by classifying the galaxies multiple times,  the neural networks perform as good in a fraction of the needed time.  However, for dwarf irregulars we are still not at a level where a label can be taken with confidence. This has two sources: they a) come in many different shapes and sizes and don't have clear defined structures, and b) are quite infrequent, meaning that the training set is sparse. Adding more dwarf irregulars to the training sample will certainly help alleviating this problem. 

Comparisons to a commonly used network in the literature -- ResNet50 -- which was pre-trained on natural images, shows that our simple network yields better results {with the resources at hand}. {This is not surprising, because ResNet50 was trained on images which are highly different than ours (i.e. images taken in daylight with high dynamical ranges and well defined structures vs. images taken during night  time with extreme low contrast and few features). However, due to the limited computing resources at hand, we only re-trained the last layer of ResNet50. A full re-training might lead to better results.}

What could be done to improve the classification? Certainly, adding more training data would result in a better accuracy, but this cannot easily be done for the data set we used. However, the ground truth could be improved with relying on not just one expert manually classifying the images, but multiple people labelling the data. 
Another avenue would be to test the impact of more elaborate data augmentation schemes, such as translations, elastic deformations and non-trivial rotations.

One may wonder what the use of the neural network is if all the galaxies have already been classified by hand. On the one hand, our findings regarding suitable architectures and hyperparameters can serve as a starting point and benchmark for training similar tasks on different data sets. On the other hand, there is the potential of transfer learning. Our neural network trained on the data from the Dark Energy Survey may be used on a different set of data, like for example the Kilo Degree Survey (KiDS). By using a pre-trained network, the training data needed to achieve similar (or better) results can go down by magnitudes, reducing the workload of manually classifying only a subset of the total data and then use the newly trained neural network for the remaining objects. Along these lines, it will be interesting to study what kind of data set we would need to use transfer learning on upcoming large-scale surveys like Euclid or LSST. 


\section*{acknowledgements} 
{We thank the referee -- Dr. Ashley Spindler -- for the rapid report and her suggestions to improve the manuscript.}
O.M. is grateful to the Swiss National Science Foundation for financial support. E.S. is thankful to the Werner Siemens Foundation  for financial support through project MIRACLE.

\bibliographystyle{aa}
\bibliography{paper}

\end{document}